\documentclass{sigchi-ext}
\usepackage[T1]{fontenc}
\usepackage{textcomp}
\usepackage[scaled=.92]{helvet} 
\usepackage{graphicx} 
\usepackage{balance}  
\usepackage{booktabs} 
\usepackage{ccicons}  
\usepackage{ragged2e} 
\usepackage[subtle]{savetrees}

\def\plaintitle{Augmenting Task Abstraction in Visualization Research with Artificial Intelligence} 
\def\emptyauthor{}
\def\plainkeywords{Information visualization; artificial intelligence; natural language processing; task abstraction.}

\title{Digital Collaborator: Augmenting Task Abstraction in Visualization Design with Artificial Intelligence}
\numberofauthors{5}

\author{%
  \alignauthor{%
    \textbf{Aditeya  Pandey*}\\
    \affaddr{Northeastern University} \\
    \affaddr{Boston, MA 02115, USA} \\
    \email{pandey.ad@husky.neu.edu} }\alignauthor{%
    \textbf{Andrea G. Parker}\\
    \affaddr{Georgia Institute of Technology}\\
    \affaddr{Atlanta, GA 30308, USA}\\
    \email{andrea@cc.gatech.edu} } \vfil \alignauthor{%
    \textbf{Yixuan Zhang*}\\
    \affaddr{Northeastern University}\\
    \affaddr{Boston, MA 02115, USA}\\ 
    \email{zhang.yixua@northeastern.edu} }\alignauthor{%
    \textbf{Michelle A. Borkin}\\
    \affaddr{Northeastern University} \\
    \affaddr{Boston, MA 02115, USA} \\
    \email{m.borkin@northeastern.edu}} \vfil \alignauthor{%
    \textbf{John A. Guerra-Gomez}\\   \affaddr{Northeastern University} \\
    \affaddr{San Jose, CA 95138, USA} \\
    \email{john.guerra@gmail.com} }
    * Authors contributed equally
}

\definecolor{linkColor}{RGB}{6,125,233}
\hypersetup{%
  pdftitle={\plaintitle},
  pdfauthor={\emptyauthor},
  pdfkeywords={\plainkeywords},
  bookmarksnumbered,
  pdfstartview={FitH},
  colorlinks,
  citecolor=black,
  filecolor=black,
  linkcolor=black,
  urlcolor=linkColor,
  breaklinks=true,
}

\begin{document}

\CopyrightYear{2020}
\setcopyright{rightsretained}
\conferenceinfo{CHI'20,}{April  25--30, 2020, Honolulu, HI, USA}
\isbn{978-1-4503-6819-3/20/04}
\doi{https://doi.org/10.1145/3334480.XXXXXXX}
\copyrightinfo{\acmcopyright}

\maketitle

\RaggedRight{} 

\begin{abstract}
In the task abstraction phase of the visualization design process, including in ``design studies'', a practitioner maps the observed domain goals to generalizable abstract tasks using visualization theory in order to better understand and address the user's needs. 
We argue that this manual task abstraction process is prone to errors due to designer biases and a lack of domain background and knowledge. 
Under these circumstances, a collaborator can help validate and provide sanity checks to visualization practitioners during this important task abstraction stage.
However, having a human collaborator is not always feasible and may be subject to the same biases and pitfalls.
In this paper, we first describe the challenges associated with task abstraction. We then propose a conceptual \emph{Digital Collaborator}---an artificial intelligence system that aims to help visualization practitioners by augmenting their ability to validate and reason about the output of task abstraction. We also discuss several practical design challenges of designing and implementing such systems.   
\end{abstract}

\keywords{\plainkeywords}

\ccsdesc[500]{Human-centered computing~Visualization}
\ccsdesc[500]{Computing methodologies~Artificial intelligence}
\ccsdesc[300]{Computing methodologies~NLP}

\printccsdesc

\section{Introduction}

Artificial intelligence (AI) has been used in the data information community to help improve design of visualizations~\cite{MLUI,vartak2017towards}. A visualization practitioner can get help from a variety of tools (e.g., Tableau, QlikView, SAS)~\cite{diamond2017data} to select proper visual encodings. However this step must be carefully considered in the context of user goals and tasks.  The visualization design process can be broadly divided into phases~\cite{5290695}, including the design study process model~\cite{Sedlmair}, performing task analysis to understand domain problems, and task abstraction that aims to recast user goals from domain-specific languages to a generalized terminology for better understanding and readability~\cite{5290695}.
Conducting task abstraction is an important but rigorous manual process that requires in-depth understanding of domain knowledge and familiarity with visualization literature~\cite{brehmermunznertask,10.1145/1168149.1168168,zhang2019idmvis}. 
For example, a biologist may be interested in results for tissue samples treated with LL-37 matching up with the ones without the peptide. A visualization researcher may translate this task to \textbf{compare} values between \textbf{two groups}~\cite{5290695}. However, to accurately perform task abstraction, a visualization practitioner must first choose the best abstraction framework and then the appropriate abstraction.  A practitioner has to keep up with the ever-growing task abstraction literature~\cite{brehmermunznertask,10.1145/1168149.1168168,zhang2019idmvis} and ensure that their personal biases that might come from previous work experiences do not affect their ability to perform task abstraction.

\marginpar{%
  \vspace{-280pt} \fbox{%
    \begin{minipage}{0.925\marginparwidth}
      \textbf{Task-Abstraction Frameworks} \\
      \vspace{1pc} To highlight the variation in task-abstraction methodologies, we present some distinguishing characteristics of three common frameworks: \\
      \vspace{0.5pc} \textit{A Multi-Level Typology of Abstract Visualization Tasks}~\cite{brehmermunznertask}: 
      A generic task abstraction framework that works well across disciplines and data-set types.\\
      \vspace{0.5pc} \textit{Task Taxonomy for Graph Visualization}  ~\cite{10.1145/1168149.1168168}: 
      A descriptive framework for tasks in the field specific to graph visualizations. This taxonomy provides more descriptive identification of visualization goals than a generalized framework~\cite{pandey2020cerebrovis}.\\
      \vspace{0.5pc} \textit{Hierarchical Task Abstraction~(HTA)}~\cite{zhang2019idmvis}: 
      HTA highlights the importance of integrating context and leverages existing task abstraction frameworks in combination with a systematic analysis of user tasks, goals, and processes. \\
      
    \end{minipage}}\label{sec:sidebartaxonomy} }
    

As task abstraction is a manual and subjective phase of the visualization design process, we argue that it may be prone to human-judgment errors. For example, domain experts often serve as project collaborators to help visualization researchers and practitioners validate the task analysis and abstraction in human-centered studies. 
However, it is challenging to have collaborators' involvement in many situations. Furthermore, human collaborators are still prone to pitfalls like keeping pace with recent development of task-abstraction theories and priming biases. Therefore, we propose an AI-enabled \emph{Digital Collaborator (DC)} that can serve as a feasible alternative to a human collaborator. We envision that a DC can assist visualization researchers by being up-to-date with task-abstraction frameworks to help identify the most appropriate framework for abstraction and can help validate the task analysis and abstraction process to identify judgment errors or biases. With this paper we hope to open a discussion on the advantages and challenges of building a DC for the visualization community.

\section{Challenges of Performing Task Abstraction}\label{sec:challenges}

We first discuss the main challenges associated with the task abstraction process. 

\textbf{A Wide Range of Task Abstraction Approaches:}
Visualization researchers have proposed various task abstraction approaches (e.g.,~\cite{brehmermunznertask,10.1145/1168149.1168168,zhang2019idmvis}). On page~\pageref{sec:sidebartaxonomy} (side-bar), we discuss three common visualization task abstraction frameworks and explain how they differ. Adopting an appropriate task abstraction approach is pivotal for visualization design as it impacts the choice of visualization design and interaction idioms. However, selecting a proper task abstraction framework requires an extensive comparison of existing literature.

\textbf{Interpretation of a Task Abstraction Framework:} 
Task abstraction is a subjective evaluation of the domain experts' needs. Subjective assessments are prone to errors arising from variability in the practitioner's understanding of an abstraction framework or an innate bias such as recency bias where the task-abstraction may be influenced by recent work. Such abstraction biases can lead to a ``domino'' effect of errors that can only be objectively verified after prototyping~\cite{Sedlmair}. Additionally, the analytic-task focused taxonomies require mastery of the terminology and definitions~\cite{10.1145/1168149.1168168}. For example, in network abstraction, it is common to use the term \textit{Topology} for properties related to the structure of the network. Topology is a mathematical term, and practitioners coming from design backgrounds may be unfamiliar with its meaning. 



\section{Automate Task Abstraction using AI}

\begin{marginfigure}[1pc]
  \begin{minipage}{\marginparwidth}
    \includegraphics[width=\marginparwidth]{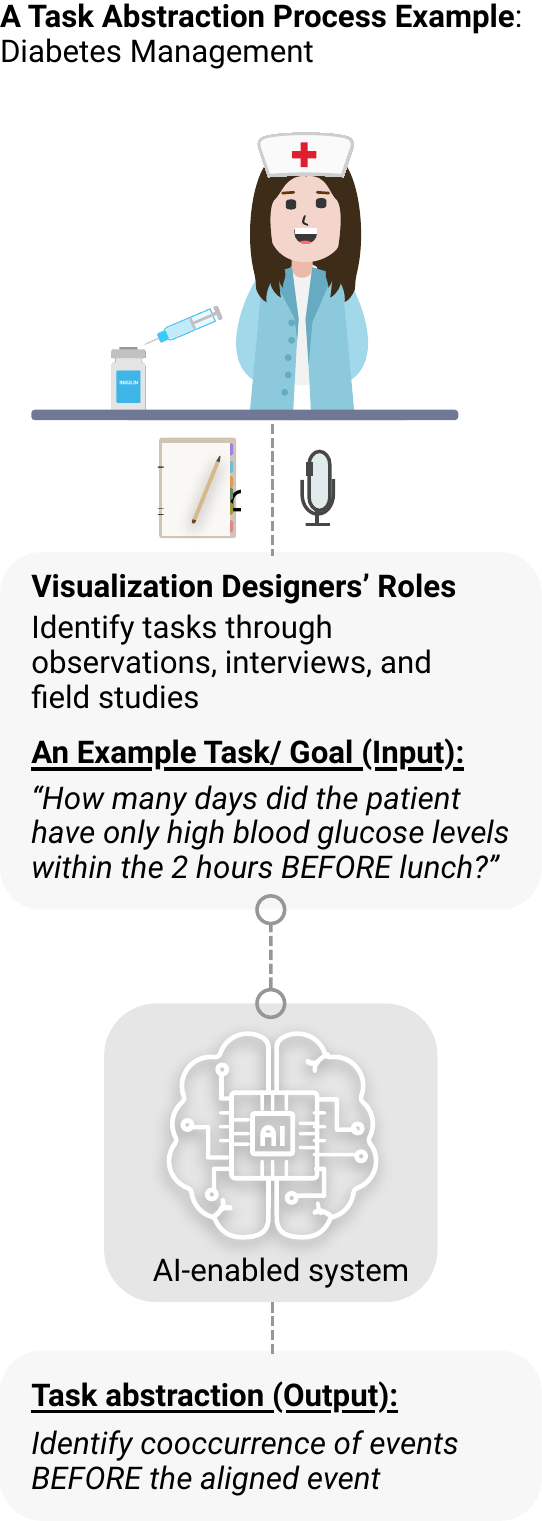}
    \caption{Example process to use AI to facilitate task abstractions in visualization research.}~\label{fig:summary}
  \end{minipage}
\end{marginfigure}

We have discussed some challenges of performing task abstraction with human effort involved. Drawing inspiration from the idiom of an Intelligent Personal Assistant (IPA)~\cite{czibula2009ipa}, we propose a Digital Collaborator (DC)---a conceptual AI-enabled system to support task abstraction for visualization research. Figure~\ref{fig:summary} shows an example of how AI can be used to facilitate task abstraction. 

\textbf{Input and Output:} Similar to the IPA system, our proposed DC will adopt a  question-and-answer-based interface. The questions (\emph{input}) will be domain goals identified by visualization practitioners through interviews and observations with domain experts. 
The DC should generate a translation of a domain goal to a generalized task description by applying an appropriate task-abstraction framework (\emph{output}). To improve communication transparency, the DC should aim to provide the rationale for their output and a set of alternative translations. 

\textbf{AI System:} 
We believe system goals should include identifying the right abstraction framework and recommending the appropriate analytical conversion of the tasks.
Note that we do not intend to suggest replacing the human-centered approaches when conducting task analysis (e.g., field studies, interviews, and observations). Instead, we propose to leverage AI in designing systems to help ease the process of task abstraction.

\section{Challenges of Our Proposed AI-enabled System}
In order to develop an AI to help automate task abstraction we acknowledge that there will be challenges to design such an AI-enabled DC system. 

\textbf{Framework Characterization:}
Task-Abstraction frameworks are well established. However, there is little guidance on how to select the ``right'' framework. 
Therefore, the first challenge of building a DC will be to develop parameters to distinguish between these abstraction frameworks.

\textbf{Training Data:} For automating the task-abstraction process, we need to train machine learning models with task data and their labeled outputs. One way to acquire training data is by parsing domain goals and their abstractions from existing literature. Smart data crawling tools may facilitate the process of extracting tasks from research papers with little manual effort. However, even after deploying web-crawlers, there might be problems with data quality. For instance, there might be conflicting abstractions where similar tasks have different abstractions.  To counter the problem, we can think of human-in-the-loop methodology where visualization researchers working on the project can address quality issues. 


\textbf{Recommendation Validation:}
Task abstraction involves subjective evaluation and characterization of domain problems. 
Practitioners may disagree with the suggested results generated by the DC. Therefore, an open question is how to instill confidence in visualization practitioners to consider the suggested results before discarding them. For example, extending the recommendation list with confidence scores may increase transparency. Future research should examine design recommendations that can boost confidence in communicating results.   

\textbf{Equity Issues with AI:} 
There has been an increasing body of research on equity issues in AI research, such as biased datasets and algorithm transparency~\cite{osoba2017intelligence}. 
An open research question is \textit{how can an AI-based DC system promote equity}? There is a vital need for future research to examine how such systems can be made more accessible for a wide audience. For example, how a variety of voices and experiences can be captured using such systems? How nuanced aspects of these experiences contribute to different design requirements and task abstractions, which will ultimately influence the design choices? Therefore, more research is needed to further explore these issues.    

\section{Conclusion}
In this paper we propose a conceptual AI-enabled digital collaborator to assist in performing visualization task abstraction. We discuss the advantages as well as the challenges of designing such AI-enabled systems, including training data, designing for transparent communication, as well as equity issues with AI.  Through this workshop paper, we want to initiate a discussion on the topic of how AI can assist task abstraction in visualization research and how to address these challenges. 
 
\section{Acknowledgements} 
This work was supported by NSF CISE CRII award no. 1657466.

\bibliographystyle{SIGCHI-Reference-Format}
\bibliography{sample}

\end{document}